\documentclass[final,5p,times,twocolumn]{elsarticle}
\usepackage{color, soul}
\usepackage{lineno,hyperref}
\usepackage{multirow}
\usepackage{graphicx}
\usepackage{float}
\usepackage{rotating}
\usepackage{array}
\modulolinenumbers[5]

\journal{A Scientific Press}

\begin{document}
	
	\begin{frontmatter}
		
		\title{Ca substituion instead of Sr in La$_{0.58}$Sr$_{0.4}$Co$_{0.2}$Fe$_{0.8}$O$_{3-\delta}$ as a cathode electrode for IT-SOFCs}

		\author[]{Majid Jafari}
		\ead{maj.id.jef@gmail.com}
		\author[add1]{Fatemeh Yadollahi Farsani}
		\ead{F.yadollahi@ph.iut.ac.ir}
		\author[add2]{Norbert H. Menzler}
		\author[add2]{Christian Lenser}
		\author[add2]{Fabian Grimm}
		\address[add1]{Department of Physics, Isfahan University of Technology, Isfahan 84156-83111, Iran}
		\address[add2]{IEK-1, Forschungszentrum Jülich, Jülich 54306, Germany}

		\begin{abstract}
			La$_{0.58}$Ca$_{0.4}$Co$_{0.2}$Fe$_{0.8}$O$_{3-\delta}$ (L58CCF) was synthesized and evaluated as a cathode electrode for intermediate temperature solid oxide fuel cells (IT-SOFC) based on the Y$_2$O$_3$-stabilized ZrO$_2$ (YSZ) electrolyte. The effect of sintering temperature on the L58CCF performance was investigated. The best Area specific resistances (ASRs) for the L58CCF sintered at 950 $^{\circ}$C were 1.218, 0.447, 0.228, 0.156, 0.099  $\Omega.cm^{2}$  at 600, 650, 700, 750, 800 $^{\circ}$C, respectively.

		\end{abstract}
		
		
	\end{frontmatter}
	

	\section{INTRODUCTION}\label{Int}
	The development of clean energy generators especially fuel/electrolyzer cells has attracted much interest due to the depletion of fossil fuel supplies and growing worries about climate change\cite{jeerh2021recent,kalair2021role}.
	Because of their high energy conversion efficiency, direct conversion of chemical energy into electrical energy, fuel adaptability, and environmentally friendly nature, solid oxide fuel cells (SOFC) are considered essential components of the promised Carbon-free energy cycle\cite{mahmud2017challenges,damo2019solid,hwang2017perovskites}. Lowering the operating temperature and improving long-term operation, particularly for cathodes, are  ongoing topics, as electrochemical performance degradation is primarily caused by slow kinetics of the oxygen reduction reaction and weak durability under long-term working conditions\cite{yang2020toward,yin2019oxide}. As a result, it is still worthwhile to devote significant resources to the cathode development.
	
	Generally, perovskite oxides have been suitable cathode materials because the physical and chemical characteristics of ABO$ _{3} $-type perovskites could be readily adjusted by substituting various dopants for A/B-site cations.
	The most appealing cathode systems are simple and double-doped cobaltite-ferrites, which show strong electrochemical activity in cells with both oxygen- and proton-conducting electrolytes\cite{pikalova2019functionality}. Much effort has been devoted to the development of high-efficiency and stable cathode materials. The A/B-site doping of LaCoO$ _{3} $ and LaFeO$ _{3} $ with alkaline earth and transition metal has received considerable attention. For example, a higher oxygen vacancy content may be obtained by replacing La with alkaline earth metals such as Ca, Sr and Ba at the A-site, and the electrocatalytic activity of the cathode is enhanced by manipulating the B site by transition metals like Co, Fe and Ni \cite{shannon1976revised}.
	
	(La,Sr)(Co,Fe)O$_{3-\delta}$ is one of the most common cathode materials for IT-SOFC applications. For LSCF cathodes, Sr separation is considered to be the mainly possible degradation mechanism. Because Sr separation, in addition to de-activate sites for Oxygen reduction reaction, causes the chemical reaction between the LSCF cathode and ZrO$ _{2} $-based electrolytes.
	
	LCCF exhibits more stable behavior than LSCF because Ca has a closer atomic radius to La than Sr (Ca = 1.34 Å; La = 1.36 Å and Sr = 1.44 Å). In fact, the radius mismatch for Ca is minimized compared to Sr.
	
	In this work, we apply a thin barrier layer of Gadolinia-doped Ceria (GDC) on the Yttria-stabilized Zirconia (YSZ) to avoid its reaction with La$_{0.58}$Ca$_{0.4}$Co$_{0.2}$Fe$_{0.8}$O$_{3-\delta}$ (L58CCF) cathode.

	\section{EXPERIMENTAL DETAILS}\label{Exp}
	
	La$_{0.58}$Ca$_{0.4}$Co$_{0.2}$Fe$_{0.8}$O$_{3-\delta}$ (L58CCF) and was synthesized using the Pechini method and La$_{0.6}$Ca$_{0.4}$Co$_{0.2}$Fe$_{0.8}$O$_{3-\delta}$ (L6CCF) powder was synthesized by the sol-gel method\cite{jafari2019enhancement}. For L58CCF composition, stoichiometric amounts of Lanthanum nitrate hexahydrate [La(NO$_{3}$)$_{3}$.6H$_{2}$O, 99.99$ \% $, Alfa Aesar] as a La source, Calcium nitrate tetrahydrate [Alfa Aesar, 99$ \% $, Ca(NO$_{3}$)$_{2}$.4H$_{2}$O ] as a Ca source, Iron nitrate nonahydrate [Alfa Aesar, 98$ \% $, Fe(NO$_{3}$)$_{3}$.9H$_{2}$O] as a Fe source, and Cobalt nitrate hexahydrate [Alfa Aesar, 98 $ \% $ , Co(NO$_{3}$)$_{2}$.6H$_{2}$O] as a Co source were completely dissolved in water. 20$ \% $ citric acid was added to the solution as a complexing agent and stirred for 1 hour. Finally, ethylene glycol was added to the solution as a chelating agent. The products temperature was increased up to 350 $^{\circ}$C and after self-combustion, a gray powder remained. The remained ash was pre-calcined at 650 $^{\circ}$C and finally calcined at 900 $^{\circ}$C for 5 h. A pre-synthesized La$_{0.58}$Sr$_{0.4}$Co$_{0.2}$Fe$_{0.8}$O$_{3-\delta}$ (L58SCF) powder (same Pechini methode as L58CCF) in IEK-1 Forschungszentrum Jülich was used for different X-ray diffraction studies.
	
	\begin{table}[h]
		\begin{center}
			\caption{The particle size of different cathode powders}
			\label{tab1}
			\begin{tabular}{|c|c|c|c|}
				\hline
				\multirow{2}{*}{Cathode powder}&  \multicolumn{3}{|c|}{Particle size($\mu$m)}   \\	 
				{} & $ d_{10} $ & $ d_{50} $ & $ d_{90} $\\ \hline
				La$_{0.58}$Ca$_{0.4}$ Co$_{0.2}$Fe$_{0.8}$O$_{3-\delta}$ (L58CCF)& 0.60 & 0.82 & 1.10\\ \hline
				La$_{0.6}$Ca$_{0.4}$ Co$_{0.2}$Fe$_{0.8}$O$_{3-\delta}$ (L6CCF)& 0.03 & 2.92 & 5.84\\ \hline
				La$_{0.58}$Sr$_{0.4}$ Co$_{0.2}$Fe$_{0.8}$O$_{3-\delta}$ (L58SCF)& 0.55 & 0.75 & 1.00\\ \hline
			\end{tabular}				
		\end{center}
	\end{table}
	\begin{table}[h]
		\begin{center}
			\caption{Fabrication conditions of symmetric cells}
			\label{tab2}
			\begin{tabular}{|>{\centering\arraybackslash}m{1.1cm}|>{\centering\arraybackslash}m{1.6cm}|>{\centering\arraybackslash}m{1.9cm}|>{\centering\arraybackslash}m{2.6cm}|}
				\hline
				Cathode powder & Powder calcination temp. ($^{\circ}$C) & Symmetric cell sintering temp. ($^{\circ}$C) & Symmetric cell code \\ \hline
				L58CCF & 1000 & 900 & L58CCF-10-900 \\ \hline
				L58CCF & 1000 & 950 & L58CCF-10-950 \\ \hline
				L58CCF & 1000 & 1000 & L58CCF-10-1000 \\ \hline
				L58CCF & 1000 & 1040 & L58CCF-10-1040 \\ \hline
				L58CCF & 1300 & 950 & L58CCF-13-950 \\ \hline
				L6CCF & 1200 & 950 & L6CCF-12-950 \\ \hline
				L58SCF & 1080 & - & L58SCF \\ \hline
			\end{tabular}	
		\end{center}
	\end{table}

	A Laser diffraction particle size analyzer (Horiba LA 950 V2 - Retsch Tech. GmbH, Germany) was used to study the particle size of the synthesized cathode powders. The particle size of the different cathode powders are listed in Table \ref{tab1}. 
	In order to exclude any possible influences of different particle sizes, the difference between d(10), d(50), and d(90) were kept as small as possible among the three cathode powders (by wet ball milling). The focus was placed on the respective d(50) value.

	For to validate the successful synthesis, and phase analysis of the different cathode materials, a crystallographic analysis was determined by powder X-ray diffraction using the Bragg–Brentano configuration (D4 Endeavor, Bruker AXS, Cu K$\alpha$1,2 radiation, $\lambda$$\alpha$ = 1.5418 {\AA}). Also, the high-temperature XRD (HT-XRD) measurements took place in ambient air using a temperature ramp of 5 $^{\circ}$C/min. Between 100 and 1200$^{\circ}$C, diffract grams were recorded in steps of 100$^{\circ}$C.
	
	For electrode characterization, Yttria-stabilized zirconia (8YSZ, 200$\mu$m; Kerafol) electrolyte was used as the substrate for the symmetrical cells. The gadolinia-doped ceria (10 or 20 GDC) barrier layer (wet layer thickness = 35 $\mu$m) was fabricated by Screen-printing GDC on the 8YSZ electrolyte substrate and sintered at 1300 $^{\circ}$C for 5 h. LCCF58-1000 was screen-printed onto the barrier layer (GDC) and then co-sintered either at 900, 950, 1000, 1080 and 1300 $^{\circ}$C for 3 h. Symmetrical cells, LCCF/GDC/8YSZ electrolyte substrate/GDC/LCCF, were prepared in the same way; then sintered at 900 $^{\circ}$C for 3 h.

	\subsection{Electrochemical performance analysis}
	
	Electrochemical Impedance Spectroscopy (EIS) measurements on symmetric cells were conducted using an Alpha A High Performance Frequency Analyzer (Novocontrol Technologies) at a frequency range of $ 10^{6} - 10^{-1} $ Hz and an a voltage of 20 mV. Symmetric cells tests were conducted in air using a Pt plate push-contact setup utilizing Pt wire leads. For all the sample, impedance data was collected from 600-800 $^{\circ}$C.

	\section{RESULTS AND DISCUSSIONS}\label{Exp}
	
\begin{figure}[!ht]
		\begin{centering}
			\includegraphics[width=7cm,angle=0]{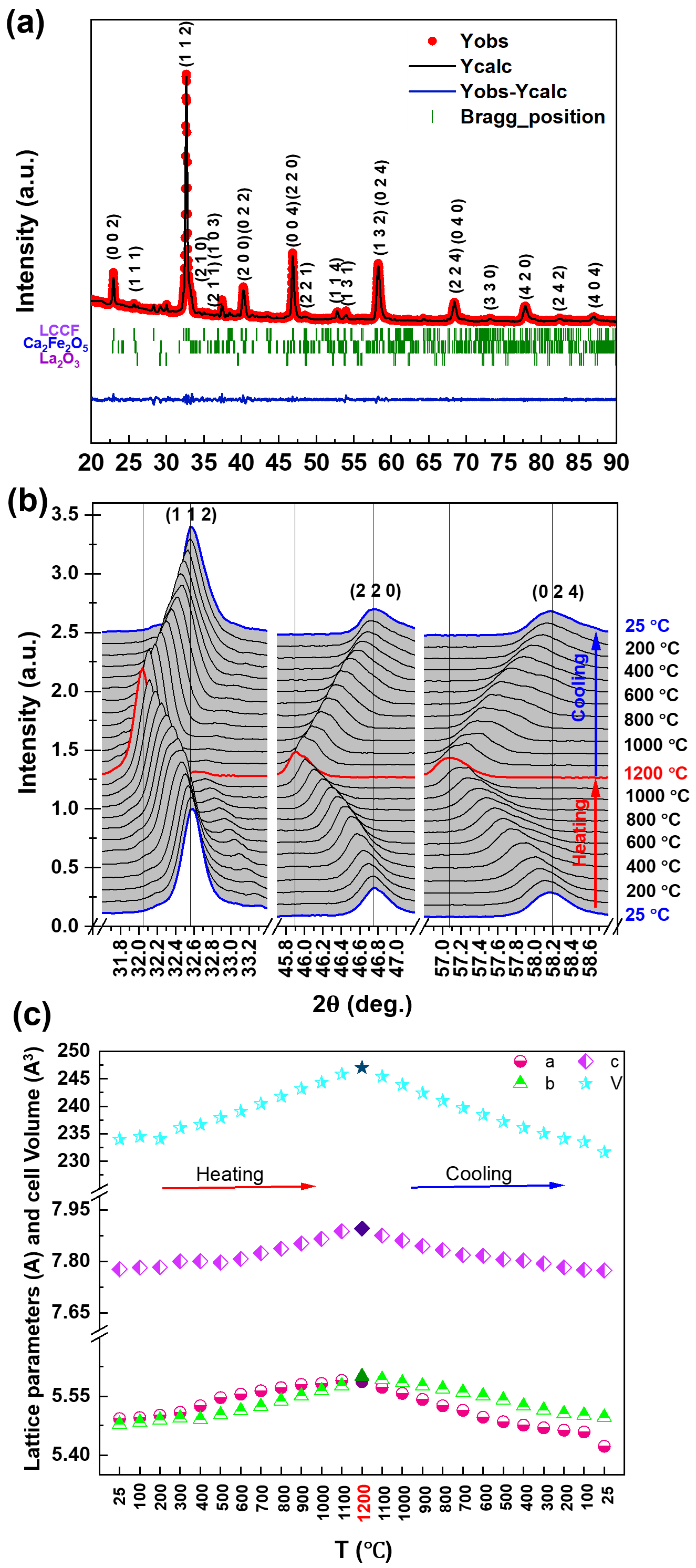}
			\vspace*{-0.4cm}
			\caption{ (a) Rietveld refinement of L58CCF-1000 at 25 $^{\circ}$C. (b) XRD pattern, (c) impurity percentage, (d) lattice parameters, and cell volume for L58CCF-1000 tested in air from 25 to 1200 $^{\circ}$C with consecutive heating/ cooling cycles.  
				\label{fig1}}
		\end{centering}
	\end{figure}

	Fig.\ref{fig1}(a) shows the crystal structure of the synthesized L58CCF powders, which were sintered in air at 1000 $^{\circ}$C for 5 h (L58CCF-1000). The obtained phase can be indexed as an orthorhombic perovskite structure with the P bnm-cab space group, in agreement with literature\cite{chen2008electrochemical,jin2008electrochemical}. The refined lattice parameter of L58CCF-1000 is a = 5.4928 {\AA}, b = 5.4787 {\AA} and c = 7.7766 {\AA} with Rp = 14.2\%, Rwp = 10.9\%, chi = 2.23.
	
	High-temperature stability of the cathode is one of the important factors for the optimization of co-sintered SOFCs, as the single cathode materials might also decompose at high temperatures leading to the formation of impurity phases. Fig.\ref{fig1}(b) shows high-temperature XRD profiles of L58CCF-1000, during heating and cooling in the temperature range 25-1200 °C. During heating and cooling, the oxide retained its perovskite structure. 
	
	The crystal symmetry and lattice parameters of the material for each temperature were determined by the Rietveld refinement. The increased (decreased) temperature causes tiny increases (decreases) in the lattice parameters without changing the orthorhombic symmetry, the P bnm-cab space group. But other impurity phases defined as La2O3, CaO and Ca2Fe2O5 were also found. These findings are supported by Efimov et al \cite{efimov2012containing} and Shen et al \cite{shen2019effect}. The La2O3 was observed at a heating rate (25-1200 ℃) and then disappeared. Also, a decrease in the amount of Ca2Fe2O5 was observed in the heating rate from about 900 °C, so that in the cooling rate after 800 °C, the amount was very small, which can be seen in Fig.\ref{fig1}(b) (the main peaks of Ca2Fe2O5 structure are indexed in the same position as those of perovskite structure \cite{salehi2017oxygen}). Also, XRD data indicated a small ~0.52° peak shift to the left and then to the right, this is likely due to an increase (decrease) in cell volume. Fig.\ref{fig1}(c) shows the changes in lattice parameters and cell volume of L58CCF-1000 powder in air with consecutive heating/cooling cycles, respectively.
	
\begin{figure}[!ht]
		\begin{center}
			\includegraphics[width=7cm,angle=0]{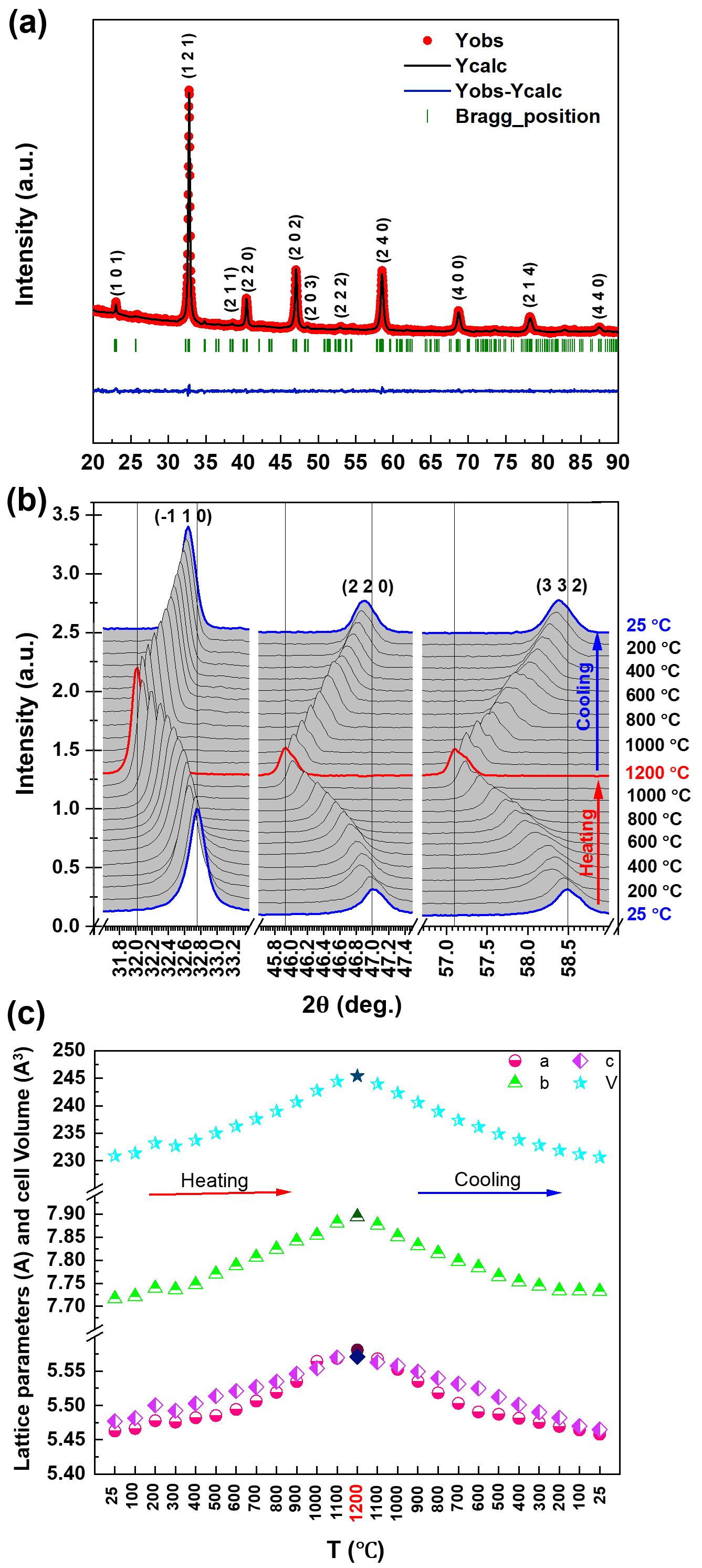}
			\vspace*{-0.4cm}
			\caption{(a) Rietveld refinement of L58CCF-1300 at 25 $^{\circ}$C (b) XRD pattern, (b) lattice parameters, and cell volume for LCCF58-1300 tested in air from 25 to 1200  $^{\circ}$C with consecutive heating/ cooling cycles.  
				\label{fig2}}
		\end{center}
	\end{figure}

	Fig.\ref{fig2}(a) shows the crystal structure of the synthesized L58CCF powders, which were sintered in air at 1300 $^{\circ}$C for 5 h (L58CCF-1300).To confirm the stability of L58CCF-1300 in the working condition, HT-XRD measurement was performed in the air during heating and cooling in the temperature range 25-1200 °C. No decomposition was observed for L58CCF-1300 in this temperature range Fig.\ref{fig2}(b), indicating that L58CCF-1300 is thermally stable. Additionally, the XRD diffraction peaks shift to a lower diffraction angle at a higher temperature for 25-1200 ℃, revealing an obvious lattice expansion at high temperatures. Such expansion is attributed to the reduction of tri- and tetra-valent Co and Fe species to a lower valence with larger ionic sizes at elevated temperatures. XRD analysis revealed an orthorhombic symmetry. The lattice parameter (P nma space group) determined by the Rietveld refinement. Fig.\ref{fig2}(c) shows the changes in lattice parameters and cell volume of LCCF58-1300 powder in air during heating and cooling in the temperature range 25-1200 °C.
	
\begin{figure}[!ht]
		\begin{center}
			\includegraphics[width=9cm,angle=0]{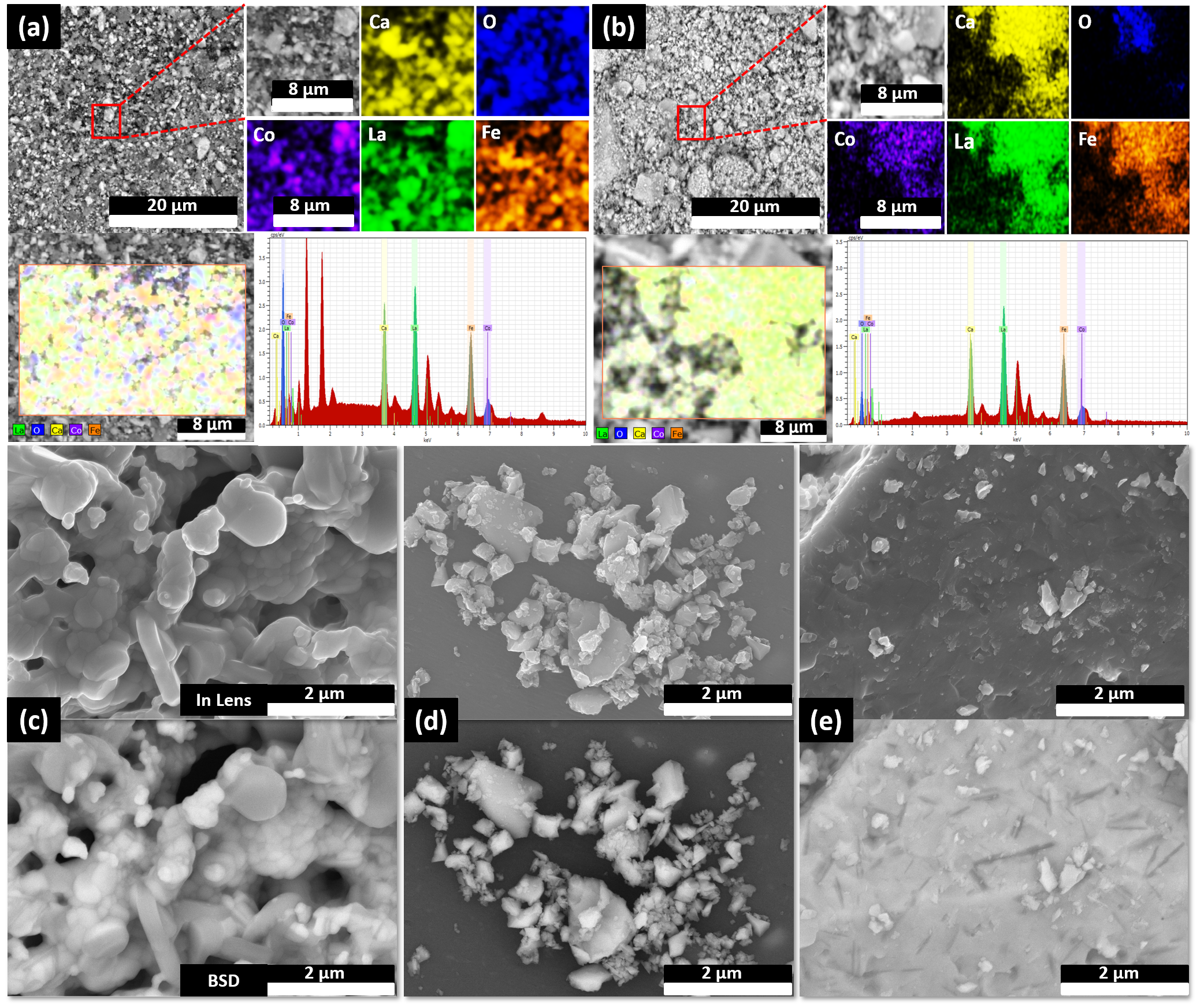}
			\vspace*{-0.7cm}
			\caption{EDX elemental distribution of La, Ca, Fe and O in the cross-sectional images for epoxy mounted and polished powders of L58CCF calcined at (a)  1000 $^{\circ}$C and (b) 1300 $^{\circ}$C.SEM micrographs of (C) L58CCF powder calcined at 1000, (d) L58CCF powder calcined at 1300 and (e) L6CCF powder calcined at 1200
				\label{fig3}}
		\end{center}
	\end{figure}

	Fig.\ref{fig3} displays SEM—backscattered electron (BSE) micrographs of the L58CCF samples calcined at 1000 ℃ and 1300 °C. According to EDX pictures, the distribution of elements, especially for the element Ca, is better for calcined powder at 1300 °C than for calcined powder at 1000 °C. However the problem is that the higher temperature leads to larger primary particle size and severer agglomeration and material need to be well milled again to small sizes.
	
	In order to study the microstructure of powders, SEM micrographs have been taken of LCFN and LCFN-YSZ-9010 composite powder samples which are presented in Fig.\ref{fig3}. A mean particle size of approximately 0.4e1 mm and also bigger agglomerations are observed for as-prepared powders which have been calcined at 800 °C for 7 h in air.
	The wet ball-milling of LCFN-YSZ-9010 for 20 h has been resulted in finer particle size distributions. The Back scattering diffraction (BSD) image of LCFN exhibited uniform distribution of formed phase without noticeable impurity or secondary different electron density phases. In the case of LCFN-YSZ-9010, the BSD image shows the presence of different phases (LCFN and YSZ), in terms of two separate phase with two different brightness at the same surface level.
	
\begin{figure}[!ht]
		\begin{center}
			\includegraphics[width=7cm,angle=0]{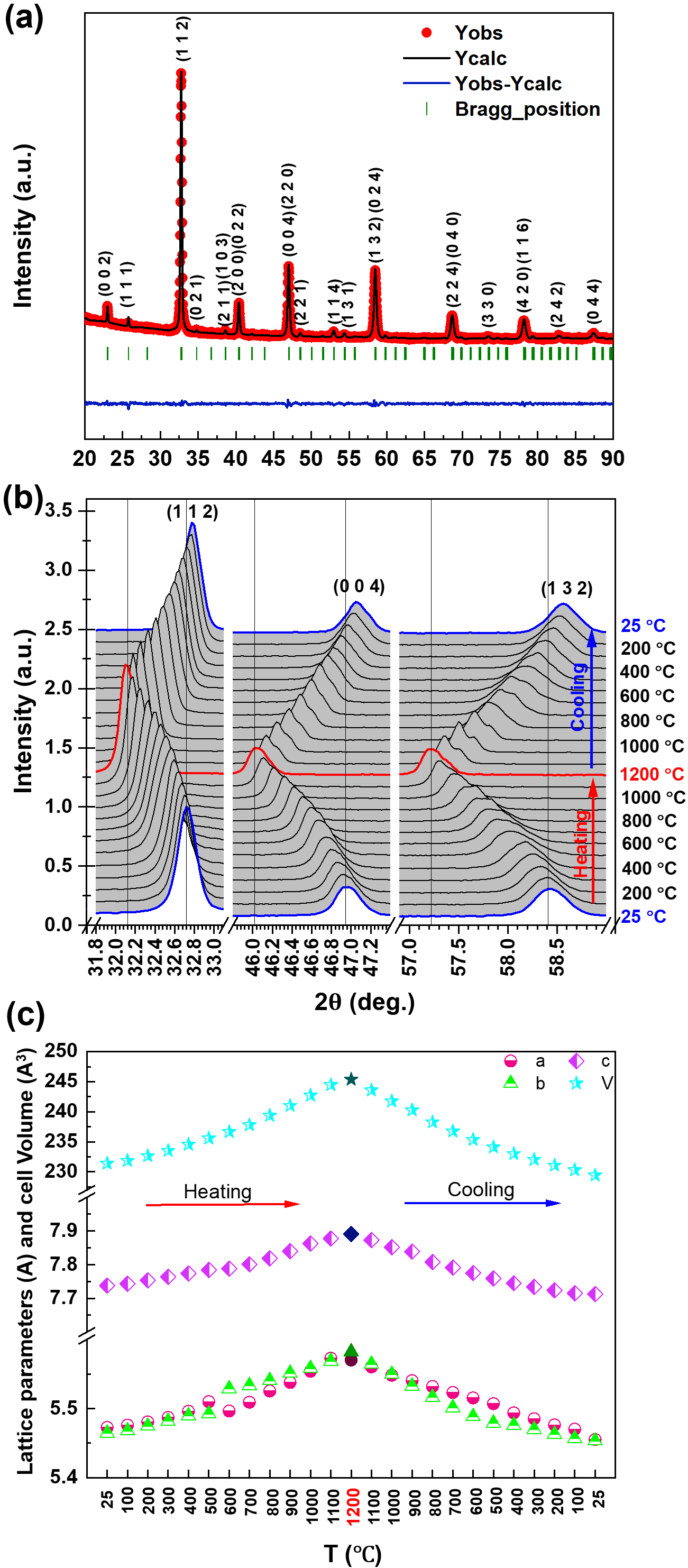}
			\vspace*{-0.4cm}
			\caption{(a) Rietveld refinement of L6CCF at 25 $^{\circ}$C. (b) XRD pattern, (c) lattice parameters, and cell volume for L6CCF tested in air from 25 to 1200  $^{\circ}$C with consecutive heating/ cooling cycles. 
				\label{fig4}}
		\end{center}
	\end{figure}

	Fig.\ref{fig4}(a) shows the Rietveld refinement of L6CCF sintered at 1200 °C from the XRD data at 25 ℃. The LCCF has a perovskite structure indexed to the orthorhombic crystal system in the P bnm space group. The Rietveld refinement calculated lattice parameters are a = 5.4727 Å, b = 5.4646 Å and c = 7.7379 Å. Also, HT-XRD patterns for L6CCF are shown in Fig.\ref{fig4}(b). As it increases from 25 ℃ to 1200 °C, L6CCF keeps the orthorhombic symmetry, with slight deviation in the peak position to a lower angle. After the temperature was decreased to 25 ℃, the deviation disappears. No secondary phases were observed for L6CCF in this temperature range. The lattice parameters and cell volume for the L6CCF determined by the Rietveld refinement. The results are shown in Fig.\ref{fig4}(C). Considering that composition with $ (A_{0.58}A^{\prime}_{0.4})B_{0.8}B^{\prime}_{0.2}O_{3} $ stoichiometry are current state-of-the-art composition \cite{grimm2020selection,han2012novel,szasz2018nature,vibhu2019high,fang2021degradation}, however, comparing the XRD results for L6CCF with the L6CCF given above, it can be concluded that the LCCF64 has better stability than the L6CCF.
	
	\begin{figure}[!ht]
		\begin{center}
			\includegraphics[width=7cm,angle=0]{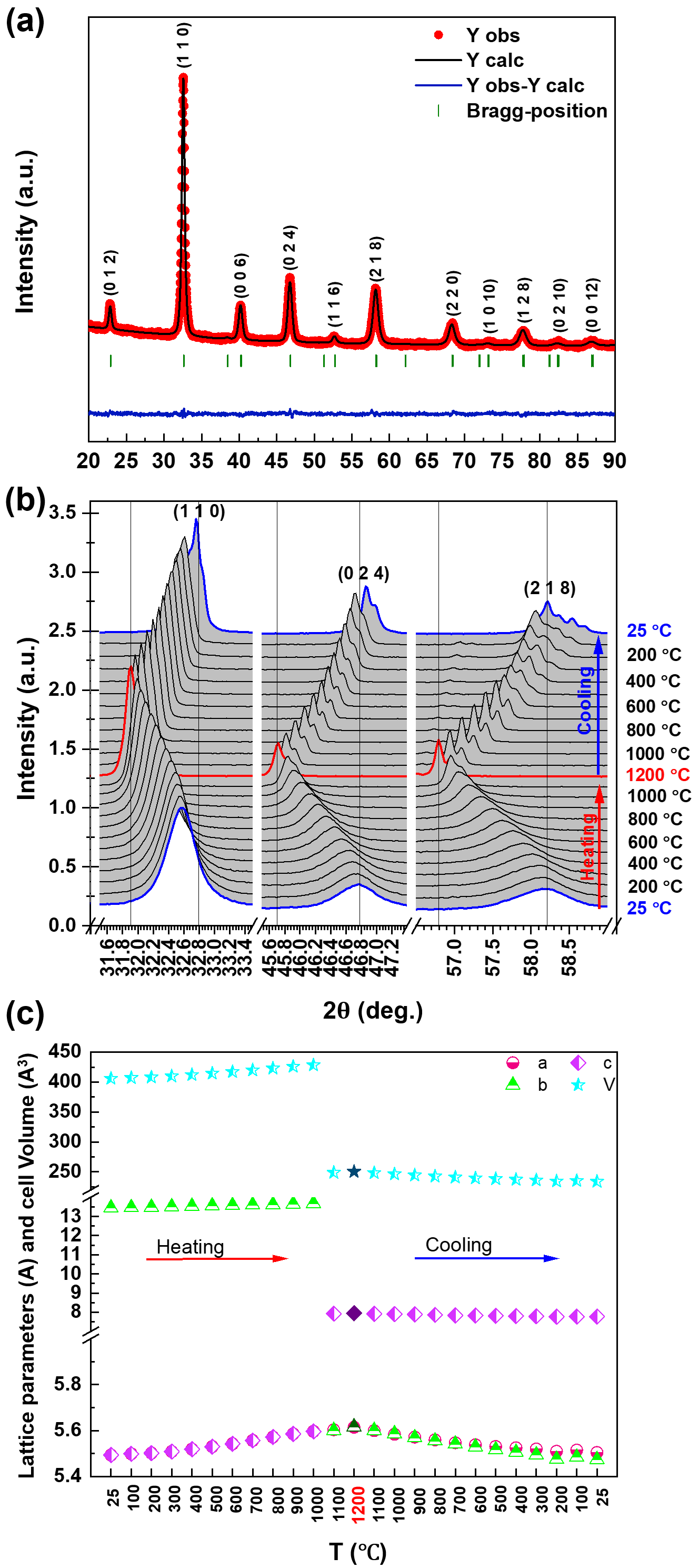}
			\vspace*{-0.4cm}
			\caption{(a) Rietveld refinement of L58SCF at 25 $^{\circ}$C. (b) XRD pattern, (c) impurity percentage, (d) lattice parameters, and cell volume for L58SCF-1080 tested in air from 25 to 1200 $^{\circ}$C with consecutive heating/ cooling cycles. 
				\label{fig5}}
		\end{center}
	\end{figure}
	
	The structural refinement of the XRD pattern of LSCF58 powder sintered at 1080 $^{\circ}$C for 3 h, was carried out by Rietveld analysis, and the results are shown in Fig.\ref{fig5}(a). It has a Rhombohedral perovskite structure with the space group R-3c.  The value of the lattice parameters obtained after the refinement were a = b = 5.4939 Å and c = 13.4500. These results are well comparable to the
	other reports \cite{vibhu2019high,celikbilek2019enhanced,pan2016study}. The stability of the LSCF58 perovskite in different temperatures was investigated using XRD technique. Fig.\ref{fig5}(b) shows the diffraction pattern of L58SFC. During heating inside the temperature range of 25–1000 °C, the L58SFC retained its Rhombohedral perovskite structure. Here further increase above 1000 °C resulted in the irreversible transition of the Rhombohedral crystal structure to the Orthorhombic crystal structure. It can be assumed that this change occurred for temperatures above 1080 ℃, which was the sintering temperature of the material. During heating, the formation of additional phases was not observed. However for cooling, small amounts of the Sr2FeO4 and CoFe2O4 impurity phases are visible. Which has been reported by Grimm et al\cite{grimm2020selection}. Fig.\ref{fig5}(C) shows the calculated lattice parameters and unit cell volumes at during heating and cooling cycles.

	\subsection{Reactivity and stability of  La$_{0.58}$M$_{0.4}$Co$_{0.2}$Fe$_{0.8}$$O_{3-\delta}$ (M = Sr or Ca) near the YSZ and GDC composite}

	For a smooth operation of a SOFC cell, it becomes important that the individual components viz., the cathode and electrolyte, show a long-term chemical stability with respect to each other. The LSCF cathode is well studied in literature in terms of its chemical stability with YSZ and GDC electrolyte systems\cite{pan2016study,chen2016direct}. However, similar studies LCCF cathode with YSZ and GDC electrolytes are unavailable to the best of our knowledge.
	
	\begin{figure}[!ht]
		\begin{centering}
			\includegraphics[width=9cm,angle=0]{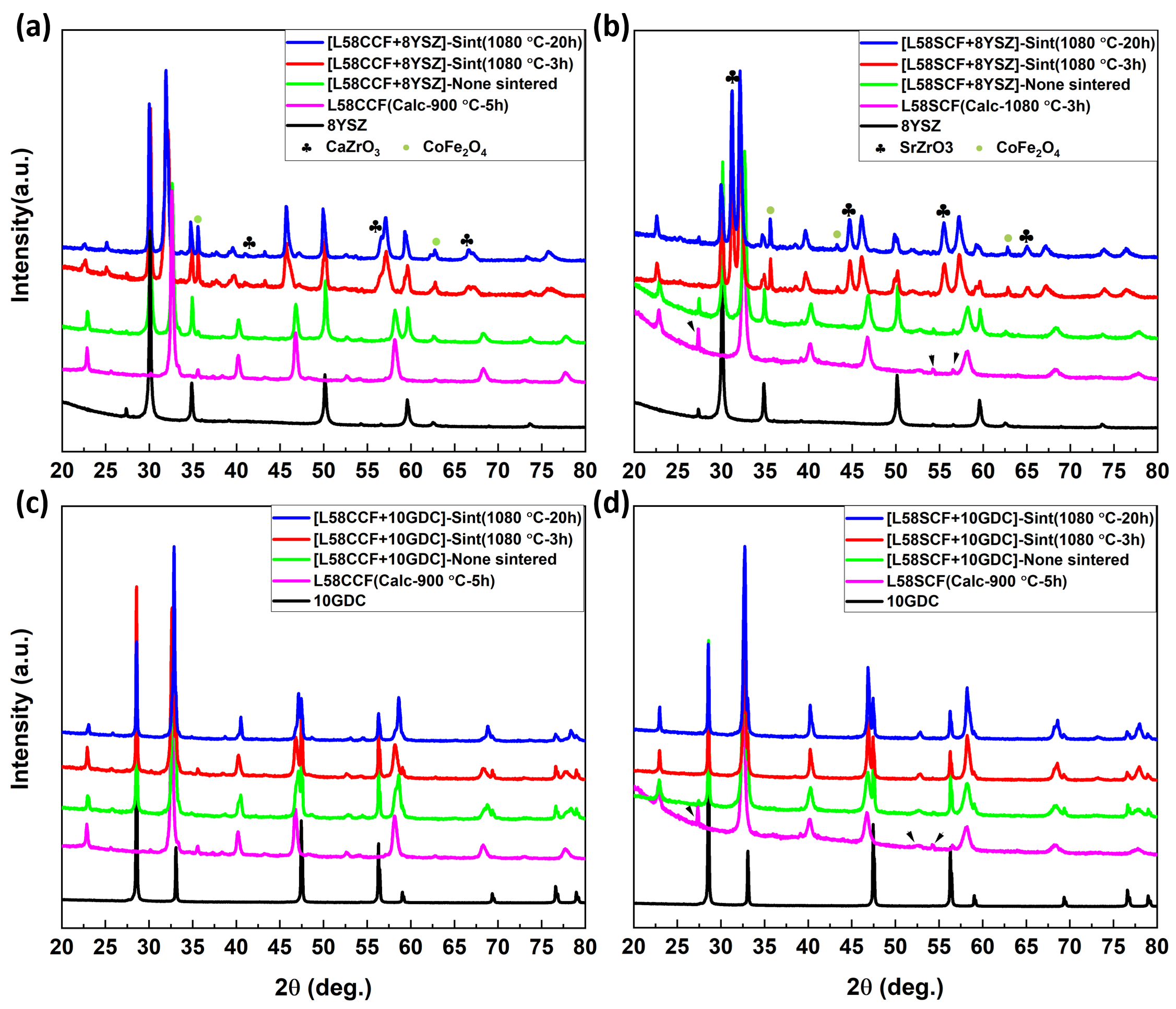}
			\vspace*{-0.7cm}
			\caption{XRD patterns for mixture of (a) L58CCF and  YSZ (58:42), (b) L58SCF and YSZ (58:42) , and mixture of (c) L58CCF and 10GDC (78-22 wt\%), (d) L58SCF and 10GDC (78-22 wt\%) sintered at 1080 $^{\circ}$C for 0 h, 3 h and 20 h in air. 
				\label{fig6}}
		\end{centering}
	\end{figure}
	
	Fig.\ref{fig6}(a) and \ref{fig6}(b) shows the X-ray diffraction patterns of (La$_{0.58}$M$_{0.4}$Fe$_{0.8}$Co$_{0.2}$O$_{3-\delta}$) (M = Sr or Ca) and YSZ (58-42 wt\%) mixture sintered at 1080 $^{\circ}$C for 0 h, 3 h and 20 h in air. For all composites, XRD data indicated a small peak shift to the left for the (La$_{0.58}$M$_{0.4}$Fe$_{0.8}$Co$_{0.2}$O$_{3-\delta}$) (M = Sr or Ca) peaks of the sintered samples as compared to the calcined powders. No shift is observed for YSZ peaks, so (La$_{0.58}$M$_{0.4}$Fe$_{0.8}$Co$_{0.2}$O$_{3-\delta}$) (M = Sr or Ca) shift is not a systematic error, but peak shifts are probably due to an increase in cell volume, which probably describes the solid state diffusion at partially melted boundaries between (La$_{0.58}$M$_{0.4}$Fe$_{0.8}$Co$_{0.2}$O$_{3-\delta}$) (M = Sr or Ca) and YSZ phases caused by diffusion of Zr4+ into the perovskite cation sublattice \cite{zink2007refractory}.

	\begin{table*}[h]
		\begin{center}
			\caption{Summary of lattice parameters extracted and impurity percentage from XRD data of the mixture of L58MCF (M=Sr and Ca) and YSZ with a weight ratio of 58:42 and 58:42 sintered at 1080 $^{\circ}$C for 0h, 3h and 20h.}
			\label{tab3}
			\begin{tabular}{|>{\centering\arraybackslash}m{1.5cm}|>{\centering\arraybackslash}m{1cm}|>{\centering\arraybackslash}m{1cm}|>{\centering\arraybackslash}m{1cm}|>{\centering\arraybackslash}m{1.1cm}|>{\centering\arraybackslash}m{1.1cm}|>{\centering\arraybackslash}m{1.7cm}|>{\centering\arraybackslash}m{0.9cm}|>{\centering\arraybackslash}m{0.9cm}|>{\centering\arraybackslash}m{0.9cm}|>{\centering\arraybackslash}m{1.1cm}|>{\centering\arraybackslash}m{1.05cm}|}
				\hline
				{Sample} & LCCF \textit{(Ortho.) a, b, c}& YSZ \textit{(Tet.) a=b, c}& CaZrO$ _{3} $ \textit{(Ortho.) a, b, c}& CoFe$ _{2} $O$ _{4} $ \textit{(Cub.)  a=b=c}& Impurity (\%)&{Sample}&L58SCF \textit{(Rhom.) a=b, c}&YSZ \textit{(Tet.) a=b, c}&SrZrO$ _{3} $ \textit{(Cub.) a=b=c}&CoFe$ _{2} $O$ _{4} $ \textit{(Cub.) a=b=c}&Impurity (\%) \\	\hline
				L58CCF-YSZ (58:42wt.\%-0h)&5.4997 5.4806 7.7766&3.6374 5.1444& - & - &  - &L58SCF-YSZ (58:42wt.\%-0h)&5.4950 13.4750&3.6374 5.1444& - & - &  - \\ \hline
				L58CCF-YSZ (58:42wt.\%-3h)&5.4854 5.5507 7.8552&3.5903 5.0100&5.5831 8.0070 5.7590&8.2639&11.46&L58SCF-YSZ (58:42wt.\%-3h)&5.5795 13.6508&3.6561 5.1626&4.0588&8.3742&36.23  \\ \hline
				L58CCF-YSZ (58:42wt.\%-20h)&5.5753 5.6618 7.9442&3.6522 5.1719&5.5831 8.0070 5.7590&8.2834&16.34&L58SCF-YSZ (58:42wt.\%-20h)&5.5897 13.6612&3.6522 5.1626&4.0588&8.3742&38  \\ \hline
			\end{tabular}	
		\end{center}
	\end{table*}

	Fig.\ref{fig6}(a) shows XRD analysis of the L58CCF-YSZ mixture powder, indicating that L58CCF reacts with YSZ, and impurity phases $ CaZrO_{3} $ and $ CoFe_{2}O_{4} $ spinel are formed. Fig.\ref{fig6}(b) shows the presence of SrZrO3 impurities, and in addition, the $ CoFe_{2}O_{4} $ spinel is formed due to the loss of the Sr element. However, forming phases of  $ SrZrO_{3} $ and $ CoFe_{2}O_{4} $ are reported by Pan et al. between L6SCF and YSZ at different temperatures \cite{pan2016study}. For the prepared samples of L58CCF and L58SCF, lattice parameters of the presented phases and the percentage of reaction between La$_{0.58}$M$_{0.4}$Fe$_{0.8}$Co$_{0.2}$O$_{3-\delta}$ (M = Sr or Ca) and YSZ are extracted from XRD data and listed in Table \ref{tab3}, respectively. The results indicate that the amount of undesired reaction with YSZ for the LCCF58 perovskites is less than that for the LSCF58 perovskites.
	
	The chemical compatibility with GDC electrolytes was investigated by heating La$_{0.58}$M$_{0.4}$Fe$_{0.8}$Co$_{0.2}$O$_{3-\delta}$ (M = Sr or Ca)-10GDC (78:22 wt\%) composites at 1080 $^{\circ}$C for 0h, 3h and 20h is shown in Fig.\ref{fig6}. In all cases, the La$_{0.58}$M$_{0.4}$Fe$_{0.8}$Co$_{0.2}$O$_{3-\delta}$ (M = Sr or Ca) and the used electrolytes (10GDC) maintain their crystalline structures without any degradation. Every peak in the pattern can be attributed to either GDC or La0.58M0.4Fe0.8Co0.2O3-$\delta$ (M = Sr or Ca), and no impurity phases are observed, indicating great chemical compatibility between the electrolyte (10GDC) and La$_{0.58}$M$_{0.4}$Fe$_{0.8}$Co$_{0.2}$O$_{3-\delta}$ (M = Sr or Ca).
	
	\subsection{Electrochemical behavior of the L58CCF electrodes in symmetrical cells}
	
	\begin{figure}[!ht]
		\begin{center}
			\includegraphics[width=9cm,angle=0]{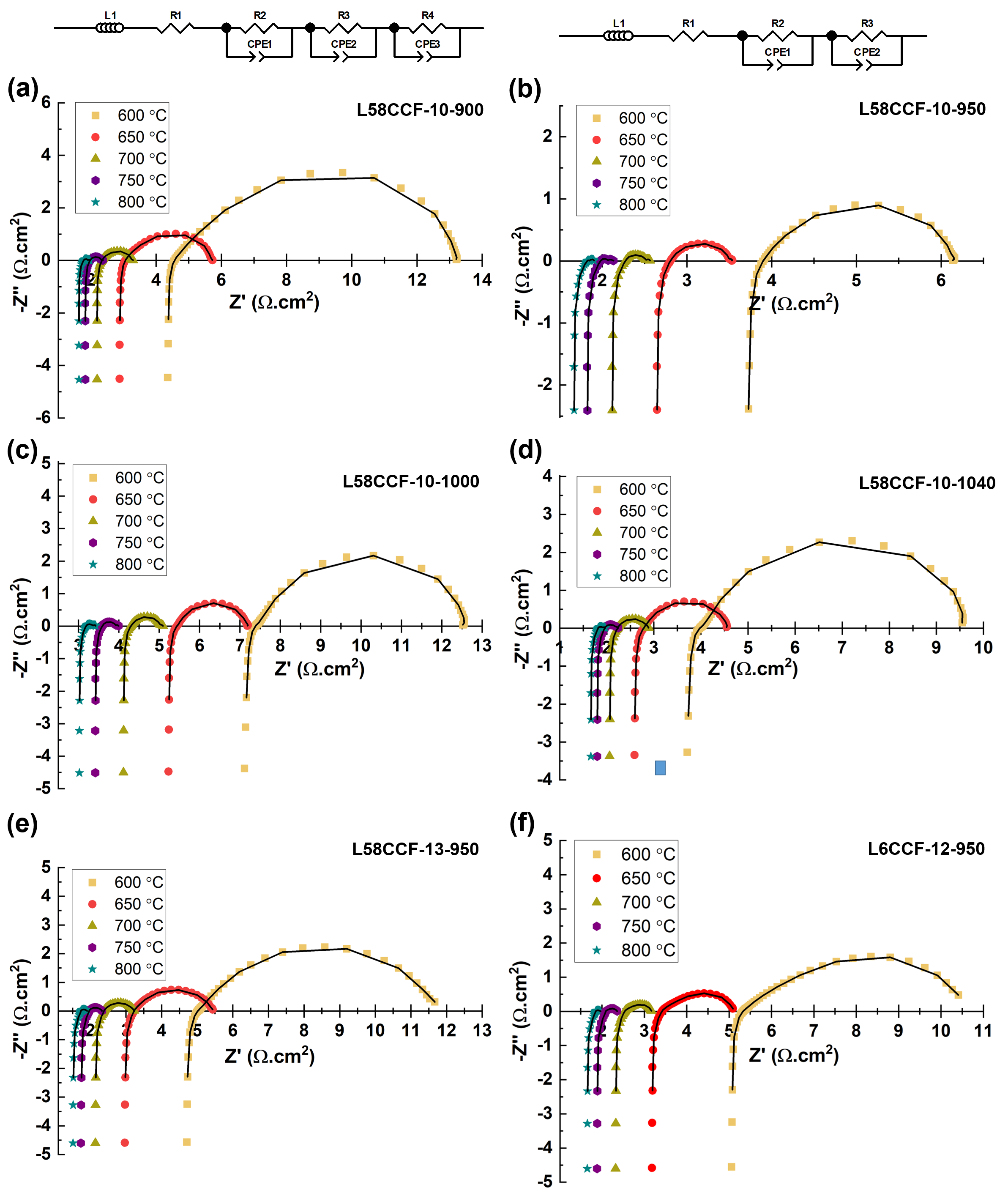}
			\vspace*{-0.5cm}
			\caption{EIS of (a) L58CCF-10-900, (b) L58CCF-10-950, (c) L58CCF-10-1000, (d) L58CCF-10-1040, (e) L58CCF-13-950, (f) L6CCF-12-950 and equivalent circuit model for symmetrical cells. 
				\label{fig7}}
		\end{center}
	\end{figure}
	
	The properties of cathode materials depend strongly on their microstructure. The microstructure also depends on the sintering temperature (TS). Low sintering temperature causes poor cathode adhesion to the electrolyte and thus increases the amount of polarization resistance. High sintering temperature, on the other hand, causes the cathode to break and crack, reducing porosity and thus increasing polarization resistance \cite{mrozinski2019electrochemical,muhammed2018enhanced,boukamp2011impedance}. In this section, the effects of the L58CCF electrode sintering temperature on the electrochemical performance were evaluated using symmetrically designed cells. Finally, the impedance spectra of L58CCF were compared with those of L6CCF cathode.
	
	Electrochemical impedance spectroscopy (EIS) plots for L58CCF/GDC—YSZ—GDC/L58CCF sintered at different temperatures (900, 950, 1000, 1040 $^{\circ}$C) and measured at different temperatures (600 $^{\circ}$C-800 $^{\circ}$C) are shown in Fig.  \ref{fig7}.
	Additionally, impedance spectra for L6CCF—GDC—YSZ—GDC—L6CCF sintered at $^{\circ}$ C (measured at 600 ℃ - 800 ℃) are shown in Fig.\ref{fig7}(f) together with the results of the fitting. The EIS spectra can be well fitted by the equivalent circuit  (L1-R1-(R2CPE1)-(R3CPE2)-(R4CPE3)) and (L1-R1-(R2CPE1)-(R3CPE2)) in ZView as shown in Fig.\ref{fig7}. L1 refers to the inductance resulting from instrumentation and ranges between 1.4 $ 10^{–6} $ H and $ 1.7 10^{–6} $ for all measurements. R1 is the ohmic resistance, which arises from the electron transport resistance in the electrode, the ion migration resistance through the YSZ and GDC electrolyte, the contact resistance between cell components, and the resistance of lead wires. CPE1, CPE2 and CPE3 are the constant phase elements for high-, intermediate-, and low-frequencies. R1, R2 and R3 are the electrode polarization resistances at high-, intermediate-, and low-frequencies, respectively. In general, R2CPE1, R3CPE2, and R4CPE3 are related to the charge transfer, surface diffusion and reactions, and gas diffusion, respectively  \cite{cui2021syngas,pichot2020electrochemical}. 
	
	\begin{figure}[!ht]
		\begin{center}
			\includegraphics[width=6cm,angle=0]{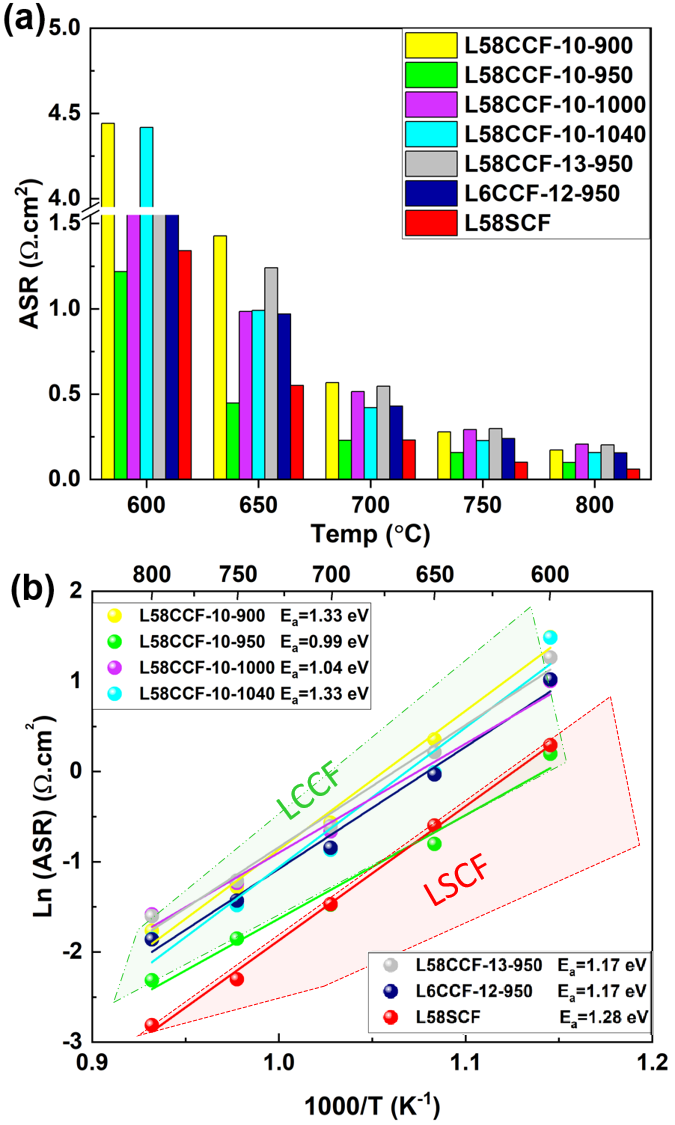}
			\vspace*{-0.3cm}
			\caption{(a) Bar chart with the respective ASR values (b) ASR Arrhenius plot of each sample in the temperature range 600 $^{\circ}$C-800 $^{\circ}$C in air.
				\label{fig8}}
		\end{center}
	\end{figure}

	The polarization resistance (RP) can be obtained by calculating the difference between the high and low frequency intercepts on the real axis (adding the three resistances R2, R3, and R4). The performance of each sample was measured in terms of the active surface area (ASR = RP/2). The values obtained for each sample (L58CCF, L6CCF, and L58SCF) in the temperature range 600 °C - 800 °C are reported in  Fig.\ref{fig8}(a). Fig.\ref{fig7}(a) shows that the L58CCF-10-950 has the lowest ASR values in the operating temperature range (600 $^{\circ}$C-800 $^{\circ}$C).The results also show that L58CCF-10-950 performs better than L6CCF. Additionally, the comparison of L58CCF-10-950 and L58SCF shows that L58CCF-10-950 performed better or similarly to L58SCF at temperatures of 600-700 $^{\circ}$C while sample 2 performed better for the rest of the temperatures. For comparison, the performance of a set of electrodes depending on a wide range of technological factors is summarized in Table \ref{tab5} \cite{khoshkalam2020improving,wang2021cerium,ortiz2014electrochemical,ascolani2019study,gilev2019high}.
	
	\begin{table*}[t]
		\begin{center}
			\caption{Different SOFC cathode materials with different configurations}
			\label{tab4}
			\begin{tabular}{|>{\centering\arraybackslash}m{2.6cm}|>{\centering\arraybackslash}m{1.2cm}|>{\centering\arraybackslash}m{1.4cm}|>{\centering\arraybackslash}m{1.7cm}|>{\centering\arraybackslash}m{1cm}|>{\centering\arraybackslash}m{0.8cm}|>{\centering\arraybackslash}m{1cm}|>{\centering\arraybackslash}m{3cm}|} 
				\hline
				Cathode&Synthesis&Electrolyte&Half cell processing&Testing temp.&ASR&E$_{a}$(eV)&Equivalent circuit  \\	\hline
				LSF\cite{khoshkalam2020improving}& & GDC &Screen-printing 1050 5hrs 25um&600 650 700 750 800& 3.211 0.980 0.358 0.149 0.069 & 1.55 & LRs(Ra//Qa)(Rb//Qb) \\ \hline
				LCSFR-GDC\cite{wang2021cerium}&sol-gel 920 $^{\circ}$C  6h &LSGM&Screen-printing 900 6hrs&800& 0.11 &-& - \\ \hline
				LCFN\cite{ortiz2014electrochemical}&liquid mix&GDC&Spraying	800 $^{\circ}$C and 1000 $^{\circ}$C- 12 h   50um & 650	700	750	800	 &4.95 1.56 0.63 0.11 0.06&  &L-Rs-RQ1-RQ2-G \\ \hline
				LSCF\cite{ascolani2019study}&sol-gel 1000 $^{\circ}$C	4 h	&LSGM&spin coating 1000 1h&500 600 700&22.0  4.0  1.36&  0.5&  -  \\ \hline
				LSCF\cite{ascolani2019study}&sol-gel 1000 $^{\circ}$C	4 h	&LSGM&spin coating 1000 1h &500 600 700&6.00  0.46  0.20&1.4&  -\\ \hline
				NBMF\cite{gilev2019high}&citrate – nitrate combustion&GDC&Paint	1100 $^{\circ}$C - 1 h&700&2.2& & L-Rs-RQ1-RQ2  \\ \hline
			\end{tabular}	
		\end{center}
	\end{table*}
	
	\begin{figure}[!ht]
		\begin{center}
			\includegraphics[width=9cm,angle=0]{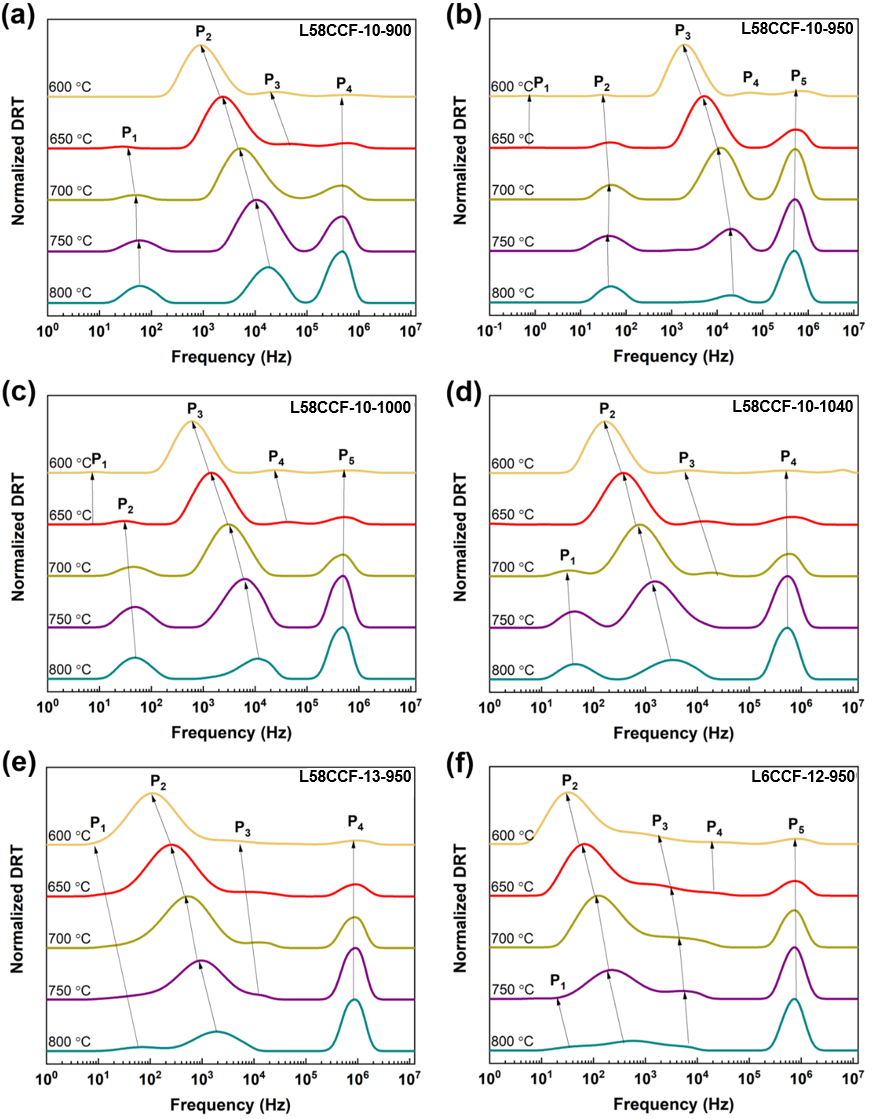}
			\vspace*{-0.7cm}
			\caption{The DRT plots of EIS spectra for (a) L58CCF-10-900, (b) L58CCF-10-950, (c) L58CCF-10-1000, (d) L58CCF-10-1040, (e) L58CCF-13-950, (f) L6CCF-12-950.
				\label{fig9}}
		\end{center}
	\end{figure}
	
	The Arrhenius plots of the ASR of the L58CCF, L6CCF and L58SCF electrode plotted as a function of temperature are shown in Fig.\ref{fig8}(b). The L58-1-950 has the lowest amount of activation energyenergy, with an activation energy of the ASR being is ~ 0.99 eV. The ASR have different activation energies.

	The Distribution of Relaxation Times (DRT) method is a relatively new approach to the analysis of impedance spectra, which allows the analysis of spectra of complex objects\cite{osinkin2021approach,gavrilyuk2017use}. A detailed analysis of the DRT functions was used to determine the partial polarization resistances. Fig.\ref{fig9} shows the result of DRT analysis using a fairly large regularization parameter of $\lambda$ = 102. According to previous reports\cite{pan2018effect,liu2021b}, the process at the frequency of around or higher than $ 10^{3} $ Hz is associated with the charge transfer (electron and oxygen ion transfer) process, the process around $ 10^{2} $ Hz represents the mass transfer (oxygen adsorption/desorption, dissociation/association, and diffusion), and the process at a frequency of around 10 Hz is related to the gas diffusion process.

	The DRT analysis of the L58CCF-10-900 revealed four relaxation processes. The P1 process with a frequency of 60-62 Hz refers to oxygen exchange and diffusion\cite{pikalova2018validation}. The P2 and P3 processes refer to charge transfer. Fig.\ref{fig9}(a) shows that the major impedance contribution (P2) was shifted to higher frequencies, and with increasing operating temperature, its share decreases. Also, the P4 process is related to inductance. For L58CCF-10-950, the P1 process is related to gas diffusion \cite{chen2017highly}, the P2 process refers to oxygen exchange and diffusion (mass transfer), the P3 and P4 processes refer to charge transfer, and the P5 process is related to inductance. At L58CCF-10-1000, the related process for each peak was the same as for L58CCF-10-950, and for L58CCF-10-1040 and L58CCF-10-1300 °C, the related process for each peak was the same as for L58CCF-10-900. It is noteworthy that for the L58CCF sample at all sinter temperatures, the main process is related to charge transfer. Fig.\ref{fig9}(f) shows the DRT analysis for L6CCF revealed five relaxation processes. The P1 and P2 processes refer to oxygen exchange and diffusion. The P3 and P4 processes refer to charge transfer, and the P5 process is related to inductance. For L6CCF, the main contribution to the polarization resistance of the cathode was related to oxygen exchange and diffusion.
	
	\begin{figure}[!ht]
		\begin{center}
			\includegraphics[width=9cm,angle=0]{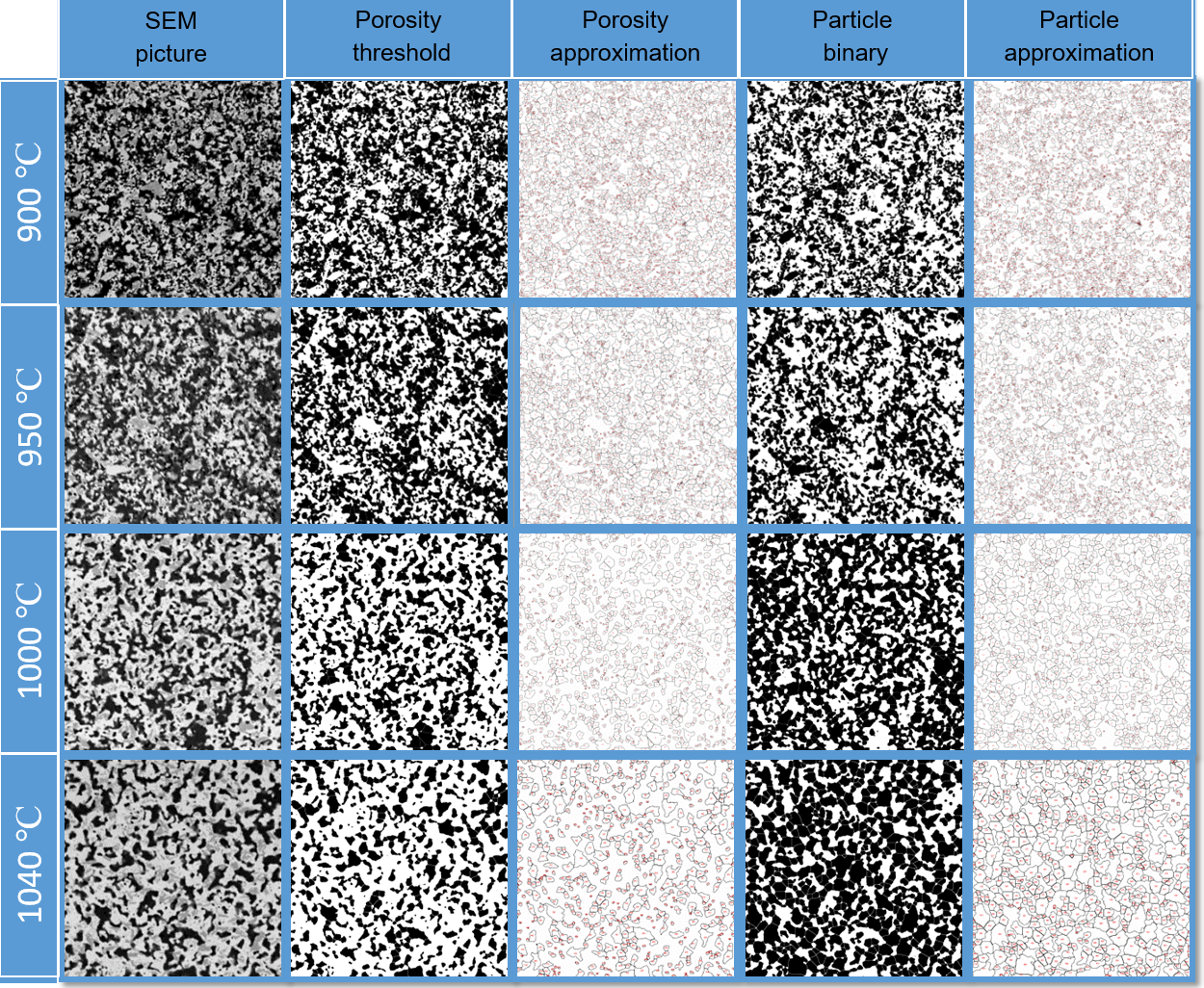}
			\vspace*{-0.7cm}
			\caption{Cross sections SEM images of L58CCF cathode film on GDC substrate with different sintering temperatures for 3 h. The ImagJ evaluations of particle sizes, average porosity sizes and their areas have been demonstrated.
				\label{fig10}}
		\end{center}
	\end{figure}
	
	\begin{table}[h]
		\begin{center}
			\caption{Average particle size and porosity obtained using ImagJ software}
			\label{tab5}
			\begin{tabular}{|>{\centering\arraybackslash}m{1.2cm}|>{\centering\arraybackslash}m{1.2cm}|>{\centering\arraybackslash}m{1.5cm}|>{\centering\arraybackslash}m{1.2cm}|>{\centering\arraybackslash}m{1.5cm}|}
				\hline
				Sintering temp.($^{\circ}$C) & Porosity area (\%) & Porosity size ($\mu$m$ ^{2} $) & Particle area (\%) & Particle size ($\mu$m$ ^{2} $) \\	\hline
				900 &53.027&0.203&44.597&0.155\\ \hline
				950  &52.403&0.310&45.701&0.257\\ \hline
				1000  &38.403&0.353&59.924&0.534\\ \hline
				1040  &33.556&0.495&64.295&0.961\\ \hline
			\end{tabular}	
		\end{center}
	\end{table}
	
	The cross-sectional SEM images in Fig.\ref{fig10} sshow the effects of sintering temperature on the grain growth of L58CCF film coated on GDC substrate. ImageJ software was used for a more detailed study. The average porosity sizes, average particle sizes, and their areas of the sintered cathode at different temperatures are listed in Table \ref{tab6}. The average particle size increased almost 6 times when the sintering temperature increased from 900 to 1040 °C. Additionally, the average porosity size increases with increasing temperature.

\begin{figure}[!ht]
		\begin{center}
			\includegraphics[width=9cm,angle=0]{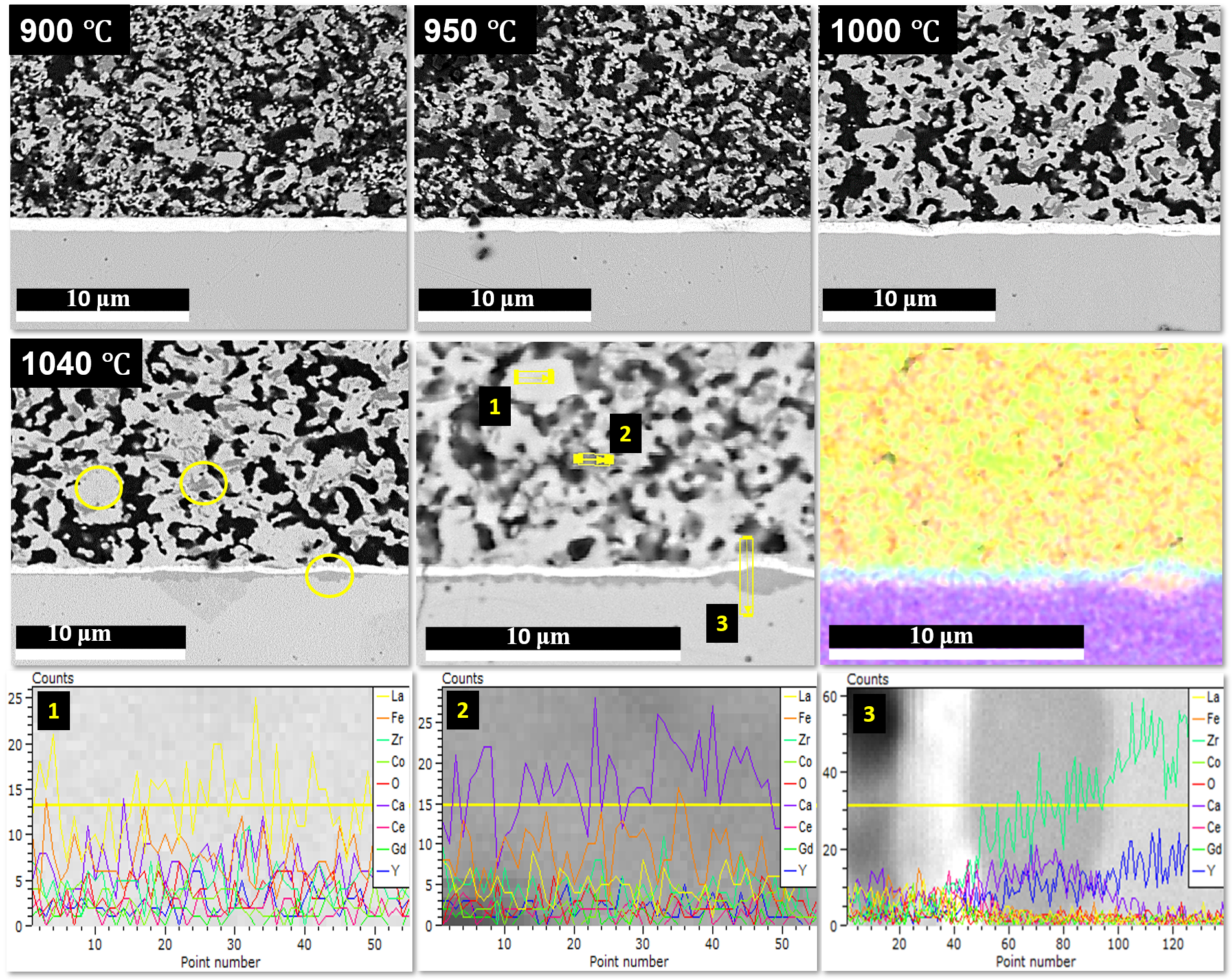}
			\vspace*{-0.7cm}
			\caption{SEM images from surfaces of the screen-printed porous L58CCF layer on top of the dense GDC electrolyte, sintered at 900 ℃, 950 ℃, 1000 ℃ and 1040 ℃ and Cross-section images of a symmetrical cell sintered at 1040 ℃ after the test with EDX line scans at three area on the cell cross-section.
				\label{fig11}}
		\end{center}
	\end{figure}

	Fig.\ref{fig11} shows post-mortem SEM images of the cross-section of symmetric cells with L58CCF electrodes sintered at different temperatures after the test. As already visible for L58CCF-10-1040, a secondary phase is formed underneath the GDC layer. For L58CCF-10-900 and L58CCF-10-950, no traces of the reaction layer were observed. A very small amount (indicated in fig.\ref{fig11}) was also observed for L58CCF-10-1000. This secondary phase was analyzed with EDX line scans. The results are shown in Fig.\ref{fig11}.  Additionally, it indicates the EDX line of arbitrarily selected areas on the surface of the cathode. As observed for the position of area 1, the EDX line confirms the presence of La, Ca, Fe, Co, and O elements. This phase is probably related to L58CCF. Besides, the distribution of elements in area 2 indicates the presence of Ca and Fe. This phase may be $ Ca_{2}Fe_{2}O_{5} $. the impurity phase of the cathode that has already been identified. The percentage of this phase is about $ 10\% $.  For the position of area 3, the formation of a secondary phase containing Ca and Zr is observed underneath the GDC layer. We guess this phase is related to $ CaZrO_{3} $. 
	
	As previously confirmed in the reaction between the electrolyte and the cathode section, the presence of this phase was confirmed. The formation of $ CaZrO_{3} $ observed in the above experiments may be due to gas-phase diffusion of Ca-containing species. Additionally, surface diffusion could also occur in parallel. Since Ca is not detected in the whole thickness of the GDC interlayer, the migration is not through bulk diffusion. F. Grimm et al. \cite{grimm2019screening} also identified this impurity phase and attributed it to Ca0.15Zr0.85O1.85. Other articles have reported that this layer is highly insulating and affects the value of RP, increasing its amount \cite{pikalova2018validation,wilde2018gd0,szasz2018nature,wankmuller2017correlative}. Z. Lu et al \cite{lu2017srzro3} reported $ SrZrO_{3} $ formation at the interlayer/electrolyte (GDC/YSZ) interface due to gas-phase diffusion of Sr-containing species. Additionally, the $ SrZrO_{3} $ formation was attributed to the pores of the SDC interlayer, and it was reported that if a dense and pore-free SDC interlayer is used, $ SrZrO_{3} $  formation is totally eliminated. J. Szasz et al. \cite{szasz2015nature},for LSCF/GDC/YSZ/GDC/LSCF symmetrical cells, reported the formation of $ SrZrO_{3} $ layer at the GDC/YSZ interface and stated that a continuous $ SrZrO_{3} $ layer is formed for GDC interlayers sintered at temperatures less than 1100°C, whereas almost no $ SrZrO_{3} $ is formed for temperatures above 1400°C. Their research revealed that the open porosity of the GDC interlayer does not decrease with increasing GDC sintering temperature.

	\section{Conclusion}
	
For L58SCF, above 1000 °C, a phase transformation from the Rhombohedral to the Orthorhombic structure occurs. In contrast, L58CCF shows no change in crystal structure at elevated temperatures. A GDC layer has been implemented to avoid cation diffusion at the interface of L58CCF/8YSZ; otherwise, a  $ SrZrO_{3} $ layer grows at intermediate SOFC operation temperatures. The reactivity studies between L58CCF and GDC revealed essentially no reaction. Among the compositions being examined, the sintered L58CCF sample at 950 °C demonstrates the minimum polarization resistance and the lowest amount of activation energy.

	\section{Acknowledgement}
	We are thankfull of Forschungszentrum Jülich for supplyig this work and providing resources. Authors would like to apreciate Dr. Sebold for FE-SEM measurements and Dr. Sohn for HT-XRD measurements. Special thanks to Torabi Shahbaz brothers (Mohammad and Mohsen) for their helpful discutions about BSD elemntal maps.

	\bibliography{mybibfile3}
	
	\bibliographystyle{elsarticle-num}

\section{supplementary}\label{Int}
Fig. 12 shows impurity percentage of L58CCF-1000, during heating and cooling in the temperature range 25-1200 °C.

\begin{figure}[!ht]
	\begin{center}
		\includegraphics[width=8cm,angle=0]{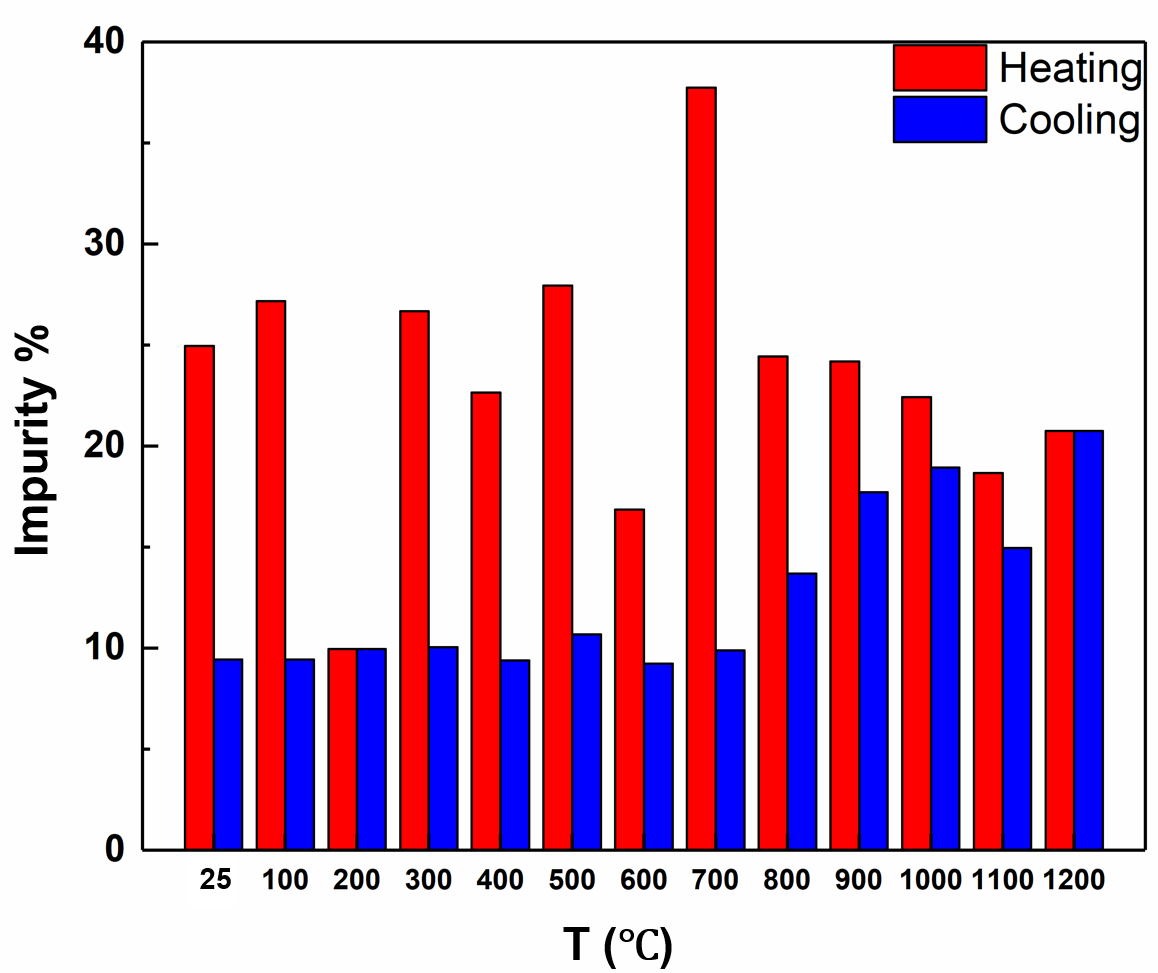}
		\vspace*{-0.4cm}
		\caption{ Impurity percentage for L58CCF-1000 tested in air from 25 to 1200 °C with consecutive heating/ cooling cycles.  
			\label{fig12}}
	\end{center}
\end{figure}

In order to eliminate impurity phases, different conditions including washing with water, different sintering temperatures, sintering times, and different atmospheres were studied. 

We washed the sample with water to eliminate impurity phases. It was observed that water-washing is somewhat effective for the elimination of the CaO and La2O3 impurity phases as indicated in Fig. 13(a). We can still see the Ca2Fe2O5 phase even after water-washing. It is observed that washing only with water is somewhat effective.

The sintering temperatures have been tested from 1000 °C to 1400 °C. As can be seen in Fig. 13(b), the impurity decreases with increasing temperature to about 1300 °C. Also at higher temperatures (1400 °C) the impurities increase and the original phase structure of the material changes.

Fig. 13(c) shows the XRD patterns of L58CCF-1000 samples prepared at 1000 ◦C for 5, 10, and 20 h in air, respectively. Increasing the sintering time had little effect on the impurity phase and the impurities remained.

Considering that the main material of L58CCF-1000 has a defective stoichiometry, it seemed that by taking oxygen from the sample and re-injecting it, the impurity could be forced to be present in the preferred lattice positions. Therefore, a small sample was annealed in argon for 5 hours at 1000 °C. A combination of two phases was again observed in the XRD pattern. Its structure was slightly different from the structure of the main material (also a change in color was observed in the sample). Of course, the formed phases had much higher peak intensities, more replaced peaks, and moved to lower angles. Therefore, the decision was made to anneal the sample again at temperatures of 900, 950, and 1000 °C in the air. The previous impurities appeared again, but it seems that they have both adjusted and that the main phase has become more pure. Baking in argon atmosphere and then in the air has rearranged the impurities. However, this method may be more useful in samples with partial impurities or for further purification of the main phase as well as the corresponding impurities. Fig. 2d shows the X-ray diffraction patterns of the L58CCF-1000 sintered in air, and argon.

\begin{figure}[!ht]
	\begin{center}
		\includegraphics[width=9cm,angle=0]{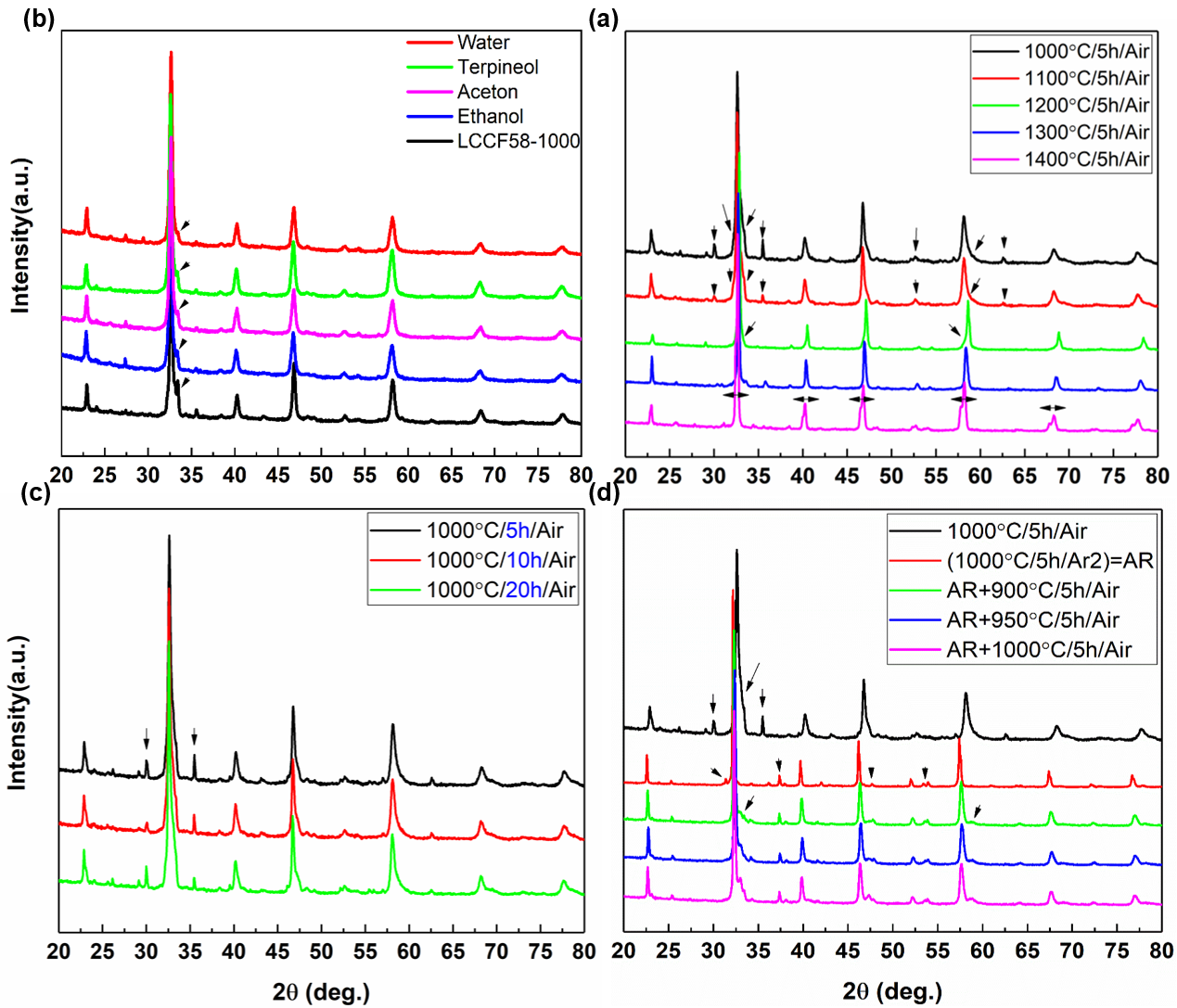}
		\vspace*{-0.4cm}
		\caption{ XRD patterns of the L58CCF composites synthesized at 1000 ◦C (a) after water-washing, (b) sintered at different temperatures, (c) for various sintering time, and (d) sintered in different atmospheres.  
			\label{fig13}}
	\end{center}
\end{figure}

According to the results, it seems that using higher temperatures for annealing of L58CCF-1000 is a more effective solution. Therefore, the prepared powders were annealed at 1300 °C for 5 h.

 Also, the results for the impurity amounts of L58SCF-1080 are shown in Fig. 14.(It is worth mentioning that due to the small amount of impurity, it was not possible to measure with maud, that is why the sum of the intensity of the impurity peaks and also the sum of the intensity of all the peaks were calculated and then the initial value was divided by the second value.)
 \begin{figure}[!ht]
 	\begin{center}
 		\includegraphics[width=8cm,angle=0]{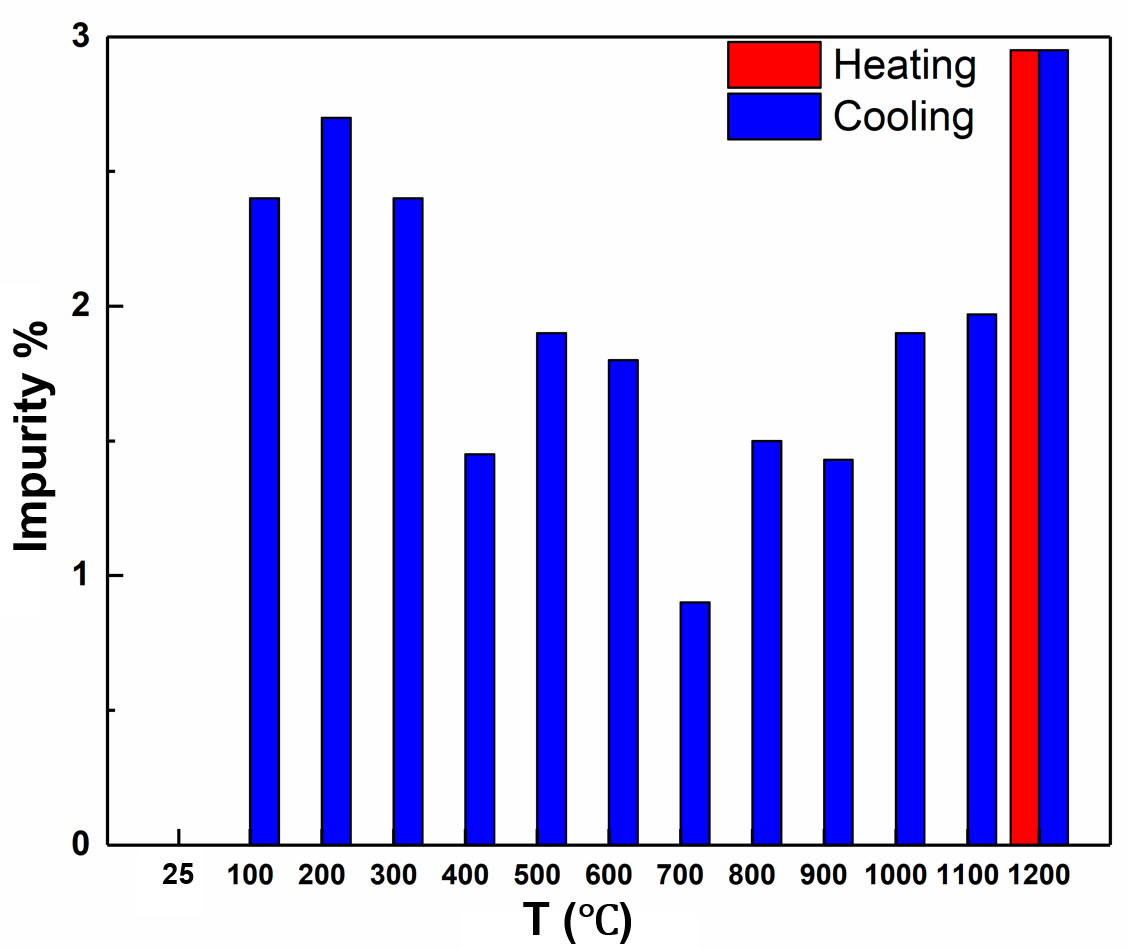}
 		\vspace*{-0.4cm}
 		\caption{ Impurity percentage for  LSCF58-1080 tested in air from 25 to 1200 °C with consecutive heating/ cooling cycles.  
 			\label{fig14}}
 	\end{center}
 \end{figure}

\subsection{Reactivity and stability of  La$_{0.58}$M$_{0.4}$Co$_{0.2}$Fe$_{0.8}$$O_{3-\delta}$ (M = Sr or Ca) near the YSZ and GDC composite}

\begin{figure}[!ht]
	\begin{center}
			\includegraphics[width=9cm,angle=0]{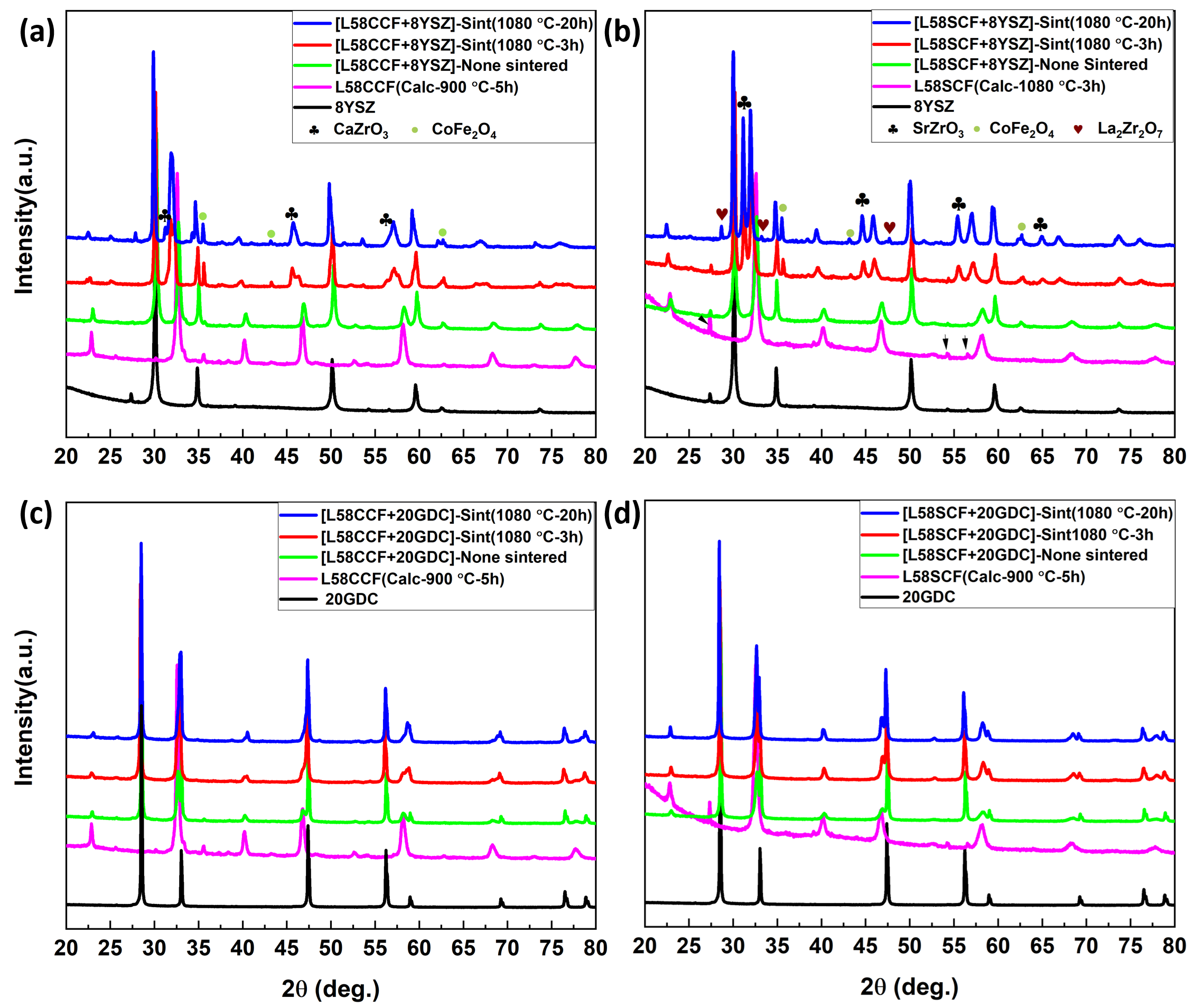}
		\vspace*{-0.7cm}
		\caption{XRD patterns for mixture of (a) L58CCF and  YSZ (50:50), (b) L58SCF and YSZ (50:50) , and mixture of (c) L58CCF and 20GDC (50-50 wt\%), (d) L58SCF and 20GDC (50-50 wt\%) sintered at 1080 $^{\circ}$C for 0 h, 3 h and 20 h in air. 
			\label{fig15}}
	\end{center}
\end{figure}

Fig.\ref{fig15}(a) and \ref{fig15}(b) shows the X-ray diffraction patterns of (La$_{0.58}$M$_{0.4}$Fe$_{0.8}$Co$_{0.2}$O$_{3-\delta}$) (M = Sr or Ca) and YSZ (50:50 wt\%) mixture sintered at 1080 ℃ for 0 h, 3 h and 20 h in air. For all composites, XRD data indicated a small peak shift to the left for the (La$_{0.58}$M$_{0.4}$Fe$_{0.8}$Co$_{0.2}$O$_{3-\delta}$) (M = Sr or Ca) peaks of the sintered samples as compared to the calcined powders. No shift is observed for YSZ peaks, so (La$_{0.58}$M$_{0.4}$Fe$_{0.8}$Co$_{0.2}$O$_{3-\delta}$) (M = Sr or Ca) shift is not a systematic error, but peak shifts are probably due to an increase in cell volume, which probably describes the solid state diffusion at partially melted boundaries between (La$_{0.58}$M$_{0.4}$Fe$_{0.8}$Co$_{0.2}$O$_{3-\delta}$) (M = Sr or Ca) and YSZ phases caused by diffusion of Zr4+ into the perovskite cation sublattice. Fig. 6(a) shows XRD analysis of the L58CFC-YSZ mixture powder shows that L58CFC reacts with YSZ and impurity phases CaZrO3 and CoFe2O4 spinel are formed. Also, no other phase like La2Zr2O7 was detected.La2Zr2O7 impurities were also observed for the L58SFC-YSZ (50:50 wt\%) mixture powder sintered for 20 h (Fig.\ref{fig15}(b)). For L58CFC and L58SFC, prepared samples, lattice parameters of the phases presented and Percentage of reaction between La$_{0.58}$M$_{0.4}$Fe$_{0.8}$Co$_{0.2}$O$_{3-\delta}$ (M = Sr or Ca) and YSZ are extracted from XRD data were listed in Table \ref{tab6}. The results show that the amount of undesired reaction with the YSZ for the L58CFC perovskites is less than the L58SFC perovskites.

\begin{table*}[h]
	\begin{center}
		\caption{Summary of lattice parameters extracted and impurity percentage from XRD data of the mixture of L58MCF (M=Sr and Ca) and YSZ with a weight ratio of 50:50 sintered at 1080 $^{\circ}$C for 0h, 3h and 20h.}
		\label{tab3}
		\begin{tabular}{|>{\centering\arraybackslash}m{1.5cm}|>{\centering\arraybackslash}m{1cm}|>{\centering\arraybackslash}m{1cm}|>{\centering\arraybackslash}m{1cm}|>{\centering\arraybackslash}m{1.1cm}|>{\centering\arraybackslash}m{1.1cm}|>{\centering\arraybackslash}m{1.7cm}|>{\centering\arraybackslash}m{0.9cm}|>{\centering\arraybackslash}m{0.9cm}|>{\centering\arraybackslash}m{0.9cm}|>{\centering\arraybackslash}m{1.1cm}|>{\centering\arraybackslash}m{1.05cm}|>{\centering\arraybackslash}m{1.05cm}|}
			\hline
			{Sample} & LCCF \textit{(Ortho.) a, b, c}& YSZ \textit{(Tet.) a=b, c}& CaZrO$ _{3} $ \textit{(Ortho.) a, b, c}& CoFe$ _{2} $O$ _{4} $ \textit{(Cub.)  a=b=c}& Impurity (\%)&{Sample}&L58SCF \textit{(Rhom.) a=b, c}&YSZ \textit{(Tet.) a=b, c}&SrZrO$ _{3} $ \textit{(Cub.) a=b=c}&CoFe$ _{2} $O$ _{4} $ \textit{(Cub.) a=b=c}&La$ _{2} $Fe$ _{2} $O$ _{7} $ \textit{(Cub.)}&Impurity (\%) \\	\hline
			L58CCF-YSZ (50:50wt.\%-0h)&5.4997 5.4806 7.7766&3.6374 5.1444& - & - &  - &L58SCF-YSZ (50:50wt.\%-0h)&5.4950 13.4750&3.6374 5.1444& - & - &  - & - \\ \hline
			L58CCF-YSZ (50:50wt.\%-3h)&5.5397 5.5853 7.8233&3.6427 5.1585&5.5860 7.9551 5.6868&8.3779&13&L58SCF-YSZ (50:50wt.\%-3h)&5.5976 13.7107&3.6458 5.1544&4.0600&8.3641&10.8076&25.96  \\ \hline
			L58CCF-YSZ (50:50wt.\%-20h)&5.6292 5.5778 7.8875&3.6556 5.1540&5.5998 8.1354 5.7646&8.3724&15.9&L58SCF-YSZ (50:50wt.\%-20h)&5.5980 13.7092&3.6468 5.1516&4.0600&8.3641&10.8076&26  \\ \hline
		\end{tabular}	
	\end{center}
\end{table*}

The chemical compatibility with GDC electrolytes was investigated by heating La$_{0.58}$M$_{0.4}$Fe$_{0.8}$Co$_{0.2}$O$_{3-\delta}$ (M = Sr or Ca)-20GDC (50:50 wt\%) composites at 1080 $^{\circ}$C for 0h, 3h and 20h is shown in Fig.\ref{fig15}. In all cases, the La$_{0.58}$M$_{0.4}$Fe$_{0.8}$Co$_{0.2}$O$_{3-\delta}$ (M = Sr or Ca) and the used electrolytes (20GDC) maintain their crystalline structures without any degradation. Every peak in the pattern can be attributed to either GDC or La0.58M0.4Fe0.8Co0.2O3-$\delta$ (M = Sr or Ca) and no impurity phases are observed, indicating great chemical compatibility between electrolyte (20GDC) and La$_{0.58}$M$_{0.4}$Fe$_{0.8}$Co$_{0.2}$O$_{3-\delta}$ (M = Sr or Ca).

\end{document}